# DDSAT: DISTRIBUTED DYNAMIC SPECTRUM ACCESS PROTOCOL IMPLEMENTATION USING GNURADIO AND USRP

Medhat H. M. Elsayed (mhamdy@qu.edu.qa)[1], Amr A. El-Sherif
(amr.elsherif@ieee.org)[2], and Amr Mohamed (amrm@qu.edu.qa)[1]
[1]affiliation: Qatar University, Doha, Qatar
[2]affiliation: Alexandria University, Alexandria, Egypt

## ABSTRACT

Frequency spectrum is one of the valuable resources in wireless communications. Using cognitive radio, spectrum efficiency will increase by making use of the spectrum holes. Dynamic Spectrum Access techniques allows secondary users to transmit on an empty channel not used by a primary user for a given time. In this paper, a Distributed Dynamic Spectrum Access based TDMA protocol (DDSAT) is designed and implemented on USRP. The proposed protocol performs two main functions: Spectrum Sensing, and Spectrum Management. Spectrum Sensing is performed to find spectrum holes in a co-operative manner using the contributing secondary users. Spectrum Management works distributively on the secondary users to allocate the spectrum holes in a fairly and efficient utilization. The DDSAT protocol is implemented using Software Defined Radio (SDR) and Universal Software Radio Peripheral (USRP). Evaluation and performance tests are conducted to show throughput and fairness of the system.

## 1. INTRODUCTION

The fixed regulation of the frequency spectrum lead to bad utilization of the spectrum. Consequently, the spectrum became non-uniform in its usage. Three regions exist: heavy use, medium use, no use at all [1]. This resulted in an inefficient spectrum utilization.

Cognitive radio (CR) techniques provide the capability of sharing the spectrum in an efficient manner. Where, spectrum holes are detected and used efficiently for other users. However, a test environment is required to test the applicability and performance of these CR techniques. The test environment incorporates PHY, MAC layers capabilities and an RF platform.

Many research work focused on implementation of a testbed for testing the MAC and CR functionalities. In [2], a software platform for building and evaluating MAC protocols and applications on dynamic spectrum access systems on USRP is proposed. However, the current version supports USRP1, which is out-of-date. In [3], the authors design and implement a hardware platform for cognitive radio applications on USRP. And in [4], the author describes the implementation of an interface between a cognitive engine and SDR software to configure the radio operation. In [5], the authors present a prototype for spectrum sensing based on MATLAB/SIMULINK with USRP. In [6], the authors present the implementation of Energy detector based spectrum sensing algorithm using GNU Radio and USRP.

Many research works have been performed to support MAC layer on an SDR platform. In [7], the authors present a complete Orthogonal Frequency Division Multiplexing (OFDM) receiver implemented in GNU Radio on USRP N210. However, this incorporates only the physical layer techniques and do not have the MAC layer timing. In [8], the authors present a PHY-MAC implementation for SDR using GNURadio and a technology called Click. Click software supports only the MAC layer functions and it does not support the PHY layer functions, thus, it has to be ported on GNURadio. In [9], a project called CROSS (Cognitive Radio Open Source Software) is implemented to allow a flexible environment for cognitive radio research. CROSS has the advantage of modularity, which facilitates the process of adding any modification in its specific function. In this work, we use CROSS system as our base model to build on.

Many research efforts have explored different MAC protocol design techniques. In [10], the authors present a testbed of an enhanced lightweight medium access control (el-MAC) protocol which performs distributed time slot allocation. And in [11], a spectrum sensing mechanism is added to the el-MAC to perform co-operative spectrum sensing.

In this paper, we present a design and implementation of a Distributed Dynamic Spectrum Access based TDMA (DDSAT) protocol using GNURADIO and USRP. The protocol performs co-operative spectrum sensing, and dynamic frequency and time slot allocation based on a priority scheduling algorithm (PSA) algorithm implemented on USRP N210.

The paper is organized as follows. Section 2 presents the system model of our work. Section 3 presents a detailed explanation of the DDSAT protocol. Section 4 discusses the system implementation and performance evaluation of the DDSAT protocol. And a conclusion is presented in section 5.

## 2. SYSTEM MODEL

We consider a wireless network that consists of secondary and primary nodes for the evaluation of our DDSAT protocol. The wireless network is implemented under the GNURadio platform. Universal Software Radio Peripheral (USRP) is used as the frontend. Each node in the network is represented by using a USRP N210 [12], which is equipped with a daughterboard SBX (400-4400) MHz Rx/Tx. The PHY and MAC processing of the DDSAT protocol is completely performed on the host computer.

In this work, we consider a wireless network that has one Base Node (BN), four Secondary Nodes (SN), and one Primary Node (PN). BN node is mainly responsible of the synchronization process in the network through the transmission of a synchronization packet every frame. Secondary nodes use the DDSAT protocol in a distributed and cooperative manner to find and allocate the spectrum holes.

**Table 1: System Parameters**

| Parameter | Value |
| --- | --- |
| CCC Frequency (Ch1) | 2.401624512 GHz |
| Channel 2 | 2.402124512 GHz |
| Channel 3 | 2.402624512 GHz |
| Channel 4 | 2.403124512 GHz |
| Modulation | GMSK |
| Sampling Rate | 0.5 MHz |

Table 1 presents the different parameters used in our system. For evaluation purposes, we only use four frequency channels and four data time slots. The BN node uses channel 1 as a Common Control Channel (CCC) for synchronization process. The remaining channel frequencies are left for the secondary and primary nodes (i.e. Spectrum sensing is performed for these 3 channels only). Time slot structure will be presented in the discussion of the protocol.

## 3. DISTRIBUTED DYNAMIC SPECTRUM ACCESS BASED TDMA PROTOCOL

In this paper, we propose a DDSAT (Distributed Dynamic Spectrum Access based-TDMA) protocol for resource allocation (i.e. frequency and time slot allocation). DDSAT is based on TDMA where time is constructed as a frame divided into a number of time slots. The DDSAT protocol performs two main processes: Spectrum Sensing, and Spectrum Management. Spectrum Sensing aims at finding the spectrum holes (e.g. Channel frequencies not currently occupied by the primary user). Spectrum Management shares the spectrum holes among the secondary users.

We propose a three state super-frame structure to perform the DDSAT protocol as shown in Figure 1. The super-frame includes Sync, DDSAT and Data states. The Sync state is only used to synchronize the current and new nodes entering the nodes, this is performed by the Base node. Each secondary node performs the protocol in the DDSAT state to sense and self allocate a channel and time slots. After allocation, each secondary node can transmit its data in the Data state. Both the DDSAT and Data states are divided into the same number of time slots (N). In our work, we confined N to 4 time slots, hence, 4 secondary nodes.

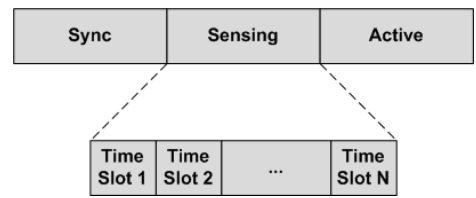

**Figure 1: Super-frame Structure for the DDSAT protocol**

Figure 2 shows the transitions between the three states in the DDSAT protocol.

**Sync State:** A separate Base node is allocated to handle the configuration of the network. The BN is responsible of performing the synchronization, time slot allocation in DDSAT state, and supporting the secondary nodes with information to use in the DDSAT protocol. This is done by periodically broadcasting a beacon packet by the BN every state (i.e. N time slots which is 4 in our work). Figure 3 shows the format of the beacon packet. Occupied time slots part of the beacon packet informs the SNs about the busy and empty time slots in the DDSAT state. List of channels part highlights the list of frequency channels to be considered in the sensing process in the DDSAT state.

A possible scenario for two secondary nodes in the network is shown in Figure 4. SN0 is a current secondary node, and SN1 is a new comer to the network. Initially, SN1 will be in the sync state waiting for the beacon packet. The beacon packet synchronizes SN1 with the TDMA frame of the DDSAT state and provides it with information about the time slot occupancy in this state. After that, it chooses a free time slot in the DDSAT state. If two secondary nodes choose the same time slot, collision will occur and the two SNs return back to the Sync State.

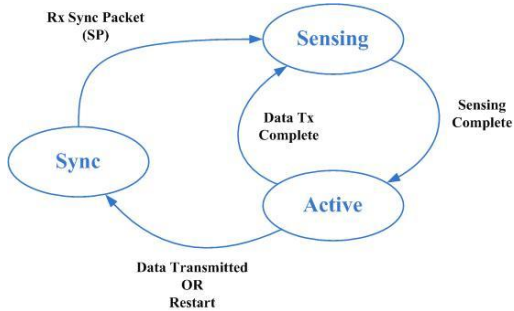

Figure 2: DDSAT States Transitions

| Packet Type | Occupied Time Slots | List of Channels |
|---|---|---|

Figure 3: Beacon Packet Format

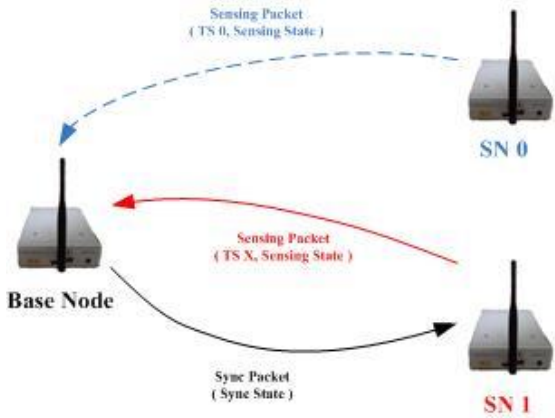

Figure 4: DDSAT Network Scenario

On the other hand, SN0 (which is a current SN) has a previous time slot allocation, thus it goes back to its previously allocated time slot in the DDSAT state (Time Slot 0, TS0).

**DDSAT State:** Following that each secondary node allocated a time slot in the DDSAT state, it begins to perform the DDSAT protocol. All secondary nodes use a Common Control Channel (CCC) to share their information for co-operative resource allocation. At the end of this state, a Priority Scheduling Algorithm (PSA) is performed independently by each secondary node to allocate channel frequency and time slot in the data state to each node.

**Data State:** Simply, each secondary node (SN) sends its data on the allocated channel frequency and time slots according to the pre-allocated time slots assigned by the PSA algorithm.

**Frame Time:** A TDMA frame is defined for both DDSAT and Data states. Same number of time slots are allocated for both states, which is 4 in this work. The slot time is 1 second plus 0.1 seconds guard interval (the long slot time is to compensate for the hardware delays in the USRP).

**Packets Format:** Two packet formats are defined: beacon packet and DDSAT packet. The beacon packet has been shown before in Figure 3, and it is used by the BN. For co-operation between secondary nodes, a DDSAT packet is used in the DDSAT where each secondary node broadcasts its information during its allocated DDSAT time slot. Figure 5 shows the DDSAT packet format.

- Empty Channels: This is a list of free channels resulted from the spectrum sensing process each SN.
- Number of Requested Slots: This is the number of requested slots by the SN.
- Priority Index: This is a priority number calculated by the PSA algorithm.
- Occupied Time Slots: The current occupied time slots. This is used to inform the BN about the time slots reserved by this SN to keep reserving it until it exits the network.
- Preferred Channels: The SN provides two preferred channels to Tx/Rx on them. This is used by the Priority Scheduling Algorithm (PSA) later.

| Packet Type | Empty Channels | Number of Requested Time Slots | Priority Index | Occupied Time Slots | Preferred Channels |
|---|---|---|---|---|---|

Figure 5: DDSAT Packet Format

The DDSAT protocol performs two main functions: Spectrum sensing and resource allocation. This is done by the secondary nodes in the DDSAT state. In the following subsections, we are going to explore the two functions to achieve high spectrum sensing accuracy, maximize secondary nodes throughput, and achieve fairness across the secondary network.

### 3.1. Sensing Algorithm

Sensing is performed using energy detection algorithm, where a threshold (T) is used to detect the primary nodes activity by each secondary node. A co-operative spectrum sensing is applied by sharing the sensing results among the secondary nodes. At the end of DDSAT state, each secondary node performs a sensing fusion to find a list of empty channel frequencies. Sensing fusion is a majority rule applied to the sensing results from the secondary nodes. We show the increase in sensing accuracy due to the co-operation in section 4.

### 3.2. Priority Scheduling Algorithm (PSA)

The PSA algorithm is used for spectrum sharing process. Two things should be shared: channel frequencies, and time slots on each channel frequency. In this work, we assume a TDMA with 4 time slots for each channel frequency.

Efficient allocation of channel frequencies and time slots requires the knowledge of the best channels for each SN and performing an algorithm to decide how many slots will be granted to that SN. The PSA on each SN is responsible of finding two preferred channels to work on, and allocating channels and time slots according to a Priority Index (PI) calculation.

**Channel Estimation:** For the sake of knowing two preferred channel frequencies for secondary node to work on, a channel estimation should be performed to get information about the channel conditions. Instead of doing channel estimation process, we use the received power as an indication of the channel conditions. Each secondary node (SN) experiences the different allocated channels in the Data state with the other nodes. In this way, nodes are learning the network and trying to find their best working parameters.

**Priority Index Calculation:** The PSA is performed separately on each SN to find the relative priorities. A Priority Index (PI) is calculated for each SN and shared between the SNs in the DDSAT packet. Equation (1) is used to calculate the Priority Index. The DT represents the Data Type and PD represents the Packet Delay. The DT takes a high value for real time data and low for normal data. The PD is a number corresponding to the delay occurred.

$$PriorityIndex = DT + PD \qquad (1)$$

The procedure of the PSA on each SN is as follows:

- The SN that has the highest PI should be allocated its best channel and the full number of the requested time slots.
- If a SN does not find enough time slots on its first preferred channel, then it switches to the second preferred channel.
- If an SN does not find enough time slots on any channels, it goes out of the frame and the PD is incremented by one to give this SN higher priority in the next time.
- If an SN does not find a free time slot in the DDSAT state, it returns back to the Sync state and the PD is incremented by one to give this SN higher priority next time.

### 4. DDSAT IMPLEMENTATION & EVALUATION

The implementation of the DDSAT protocol is performed using GNURadio software and USRP N210. We consider a wireless network that has one Base Node (BN), four Secondary Nodes (SU), and one Primary Node (PN). The TDMA frame in both the DDSAT and Data states has 4 time slots. Two evaluation tests are conducted: Functional, and Performance evaluations.

### 4.1. Functional Verification

In the functional verification, we only verify that all processes for the network is functionally correct. Such processes are spectrum sensing, preferred channel calculations, and resource allocation based on priorities.

Figure 6 presents the functional verification in case of 4 secondary nodes and one base node. After receiving the beacon packet, each secondary node selects a time slot to perform the spectrum sensing, and at the end of the DDSAT state, the sensing fusion is done. In this case, there is no primary node used, thus the sensing fusion is the three channel frequencies (2, 3, 4) as shown in the figure. Later, we present a performance evaluation test for the sensing fusion accuracy in the presence of primary node.

### 4.2. Performance Evaluation

For the performance evaluation of the DDSAT protocol, we show three performance evaluation results: throughput, fairness, and co-operative spectrum sensing accuracy.

Figure 7 shows the throughput results versus the number of secondary nodes (up to 4 secondary nodes which is the number of time slots in our configuration). In this case, perfect sensing is assumed, where a primary node occupies channel frequency 4 and the empty channels are 2 and 3. Each secondary node requests 4 time slots in the Data state to send its data. It is obvious that increasing the number of secondary nodes in the network will decrease the average throughput due to the contention on the time slots. Figure 8 shows the fairness for the same configuration.

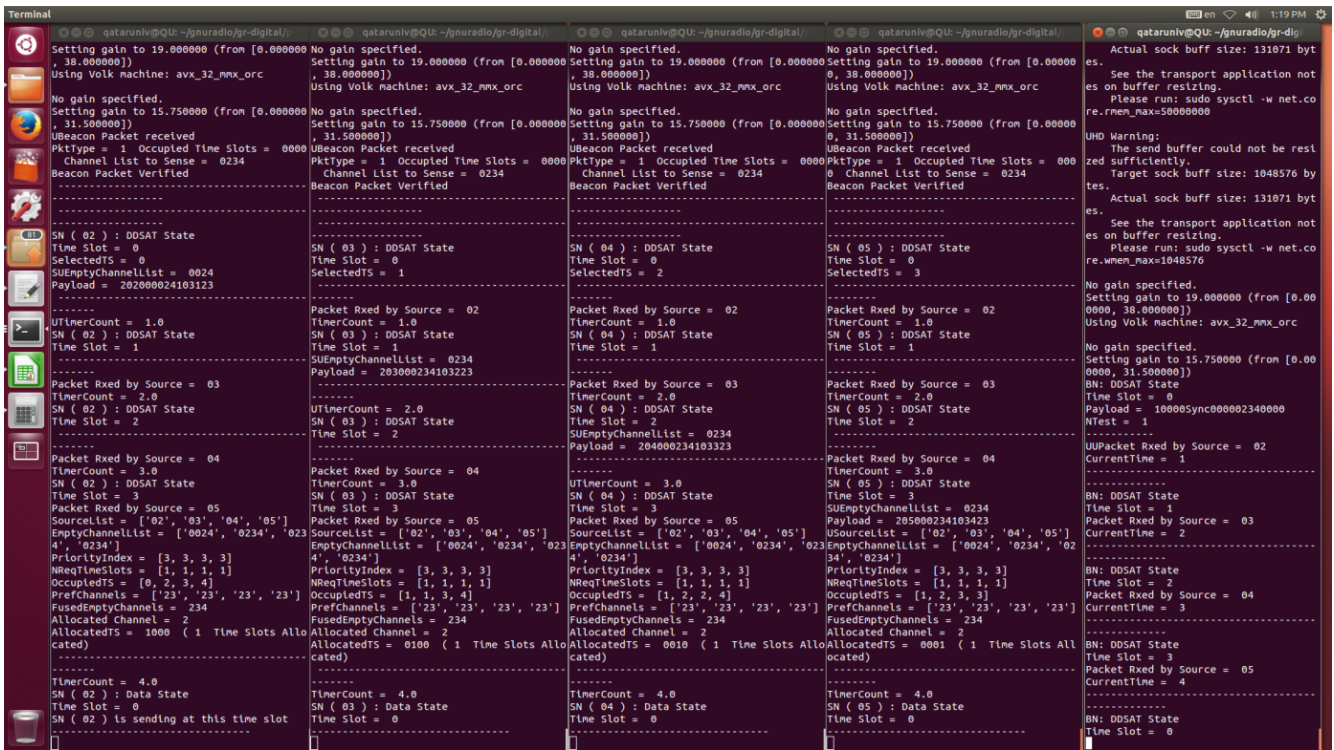

**Figure 6: Functional Verification Results with two SUs and without PU activity**

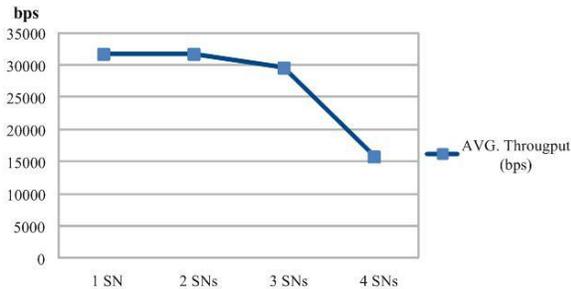

**Figure 7: Throughput (perfect spectrum sensing, maximum**

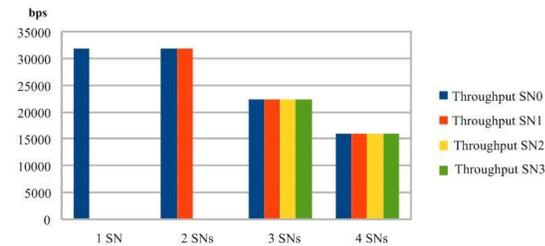

**Figure 8: Fairness (perfect spectrum sensing, maximum time**

It can be seen that the protocol is totally fair between the SNs.

Figure 9 shows the average accuracy of using the co-operative spectrum sensing. An energy detection algorithm is used to perform the spectrum sensing process, where a threshold (T) of -60 dB sustained the sensing accuracy shown in the figure. It is clear that increasing the co-operating secondary nodes will enhance the accuracy of the spectrum sensing.

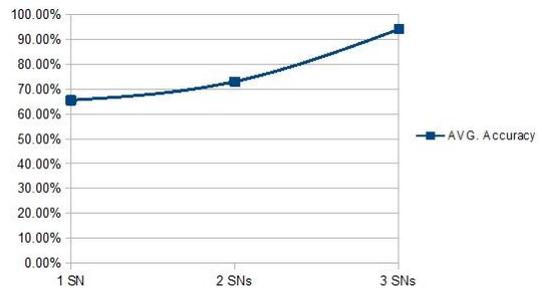

**Figure 9: Average accuracy of co-operative spectrum sensing**

## 5. ACKNOWLEDGMENT

## 6. CONCLUSION

In this work, we designed a DDSAT protocol to dynamically allocate spectrum on a secondary network. The protocol performs a distributed dynamic spectrum access on a secondary network. Spectrum sensing is performed on each SN and shared between the different SNs to improve sensing decision. Spectrum Sharing is performed through priority scheduling algorithm that calculates a relative priorities between the SNs and allocates channel and time slots based on these relative priorities. The DDSAT protocol has been implemented using GNURadio, and USRP N210 hardware platforms. Functional and performance evaluations are performed, where the protocol achieves fairness across the SNs and gives high spectrum sensing accuracy.